\documentclass[sigconf]{acmart}  
\copyrightyear{2020} 
\acmYear{2020} 
\setcopyright{acmcopyright}\acmConference[KDD '20]{Proceedings of the 26th ACM SIGKDD Conference on Knowledge Discovery and Data Mining}{August 23--27, 2020}{Virtual Event, CA, USA}
\acmBooktitle{Proceedings of the 26th ACM SIGKDD Conference on Knowledge Discovery and Data Mining (KDD '20), August 23--27, 2020, Virtual Event, CA, USA}
\acmPrice{15.00}
\acmDOI{10.1145/3394486.3403251}
\acmISBN{978-1-4503-7998-4/20/08}

\RequirePackage[rightcaption]{sidecap}
\RequirePackage{enumitem} \setlist{nosep}
\RequirePackage{soul}
\RequirePackage{bm}
\RequirePackage[textsize=scriptsize,color=yellow!25,disable]{todonotes}
\presetkeys{todonotes}{fancyline}{}
\def\framework{{\ttfamily Chatter\-Net}}

\usepackage{multirow}
\usepackage{makecell}
\RequirePackage{colortbl}
\usepackage{datetime}
\yyyymmdddate

\makeatletter
\g@addto@macro{\normalsize}{%
\setlength{\abovedisplayskip}{0pt}%
\setlength{\abovedisplayshortskip}{0pt}%
\setlength{\belowdisplayskip}{0pt}%
\setlength{\belowdisplayshortskip}{0pt}}
\makeatother
\begin{document}

\def\ztitle{Deep Exogenous and Endogenous Influence Combination for Social Chatter Intensity Prediction}

\title{\ztitle}

\author{$^1$Subhabata Dutta, $^2$Sarah Masud, $^3$Soumen Chakrabarti, $^2$Tanmoy Chakraborty}
\affiliation{%
   \institution{$^1$Jadavpur University, India; $^2$ IIIT-Delhi, India; $^3$ IIT Bombay, India}
}

\begin{abstract}
Modeling user engagement dynamics on social media has compelling applications in market trend analysis, user-persona detection, and political discourse mining.  Most existing approaches depend heavily on knowledge of the underlying user network.  However, a large number of discussions happen on platforms that either lack any reliable social network (news portal, blogs, Buzzfeed) or reveal only partially the inter-user ties (Reddit, Stackoverflow). Many approaches require observing a discussion for some considerable period before they can make useful predictions. In real-time streaming scenarios, observations incur costs.  Lastly, most models do not capture complex interactions between exogenous events (such as news articles published externally) and in-network effects (such as follow-up discussions on Reddit) to determine engagement levels. 

To address the three limitations noted above, we propose a novel framework, \framework, which, to our knowledge, is the first that can model and predict user engagement {\em without considering the underlying user network}.   Given streams of timestamped news articles and discussions, the task is to observe the streams for a short period leading up to a time horizon, then predict \emph{chatter}: the volume of discussions through a specified period after the horizon.  \framework{} processes text from news and discussions using a novel time-evolving recurrent network architecture that captures both temporal properties within news and discussions, as well as influence of news on discussions.  We report on extensive experiments using a two-month-long discussion corpus of Reddit, and a contemporaneous corpus of online news articles from the Common Crawl.  \framework{} shows considerable improvements beyond recent state-of-the-art models of engagement prediction.  Detailed studies controlling observation and prediction windows, over $43$ different subreddits, yield further useful insights.
\end{abstract}
\maketitle

\section{Introduction}
\label{sec:intro}

The Web is the most popular medium for large-scale public interaction and information propagation.  About 4.3 billion people used the Internet in 2019, with $3.53$ billion using at least one social media site\footnote{Source: https://www.statista.com/}.  Unlike radio or television, social media convey information with active participation of users. One can broadly identify two modes of engagement within user communities. In the {\em reshare} mode, a user shares some information with a community (friends, followers, groups, etc.), and members of that community recursively propagate the information. This process creates a tree of reshares, where information flows from the root to the leaves. The other is the {\em reply} mode, where one user posts some opinion, and other users reply to that post (or to other replies of the post), thus, forming a discussion. Mining the dynamics of these modes can yield useful insights for opinion mining, market research \cite{lee2015role, hu2018luxury}, political analysis \cite{jenkins2018any,ekstrom2018social} and human psychology~\cite{khan2017social}.

A growing body of research has focused on modeling the dynamics of such information propagation --- both for {\em reshare}~\citep{kobayashi2016tideh,zhao2018attentional,wang2018retweet} and {\em reply}~\citep{nishi2016reply}. There are broadly two approaches.  {\em Feature-driven models} mainly rely on three types of features for modeling the growth of reply trees, based on the social network among users, the propagated content, and temporal observations.  The other approach is to fit {\em self-exciting process} models~\citep{zhao2015seismic} (reviewed in Section~\ref{sec:Related}). In terms of growth prediction, the existing body of literature has a bias for reshare trees; specifically, Twitter retweet trees.

\paragraph{\bfseries Latent/Implicit social networks.} Most reshare models depend heavily on the user network (\emph{follow} on Twitter, \emph{friend} on Facebook), which is available on only a limited number of platforms.  However, if we focus on the dynamics of \emph{discussions}, many platforms do not offer explicit user-user social ties.  One such influential platform is Reddit.  As of \formatdate{4}{12}{2019}, it is used by $430$ million active monthly users\footnote{https://www.adweek.com/digital/reddit-reaches-430-million-monthly-active-users-looks-back-at-2019/}. Users can post {\em submissions} to one of the Reddit communities, commonly called {\em subreddits}. Other users can then {\em comment} in reply to the submission or any earlier comment on the submission. These comments form a discussion tree, with the submission at the root. Though Reddit provides a {\em subscribe} option to its users, using which they implement an internal community structure (and not an inter-user link as in Twitter or Facebook), Reddit keeps this information private.  HackerNews, IRC, and Slack assert similar constraints.  Engagement prediction methods that rely on the social network structure will lose applicability and performance.

\begin{figure}[t]
\centering
\includegraphics[width=\columnwidth]{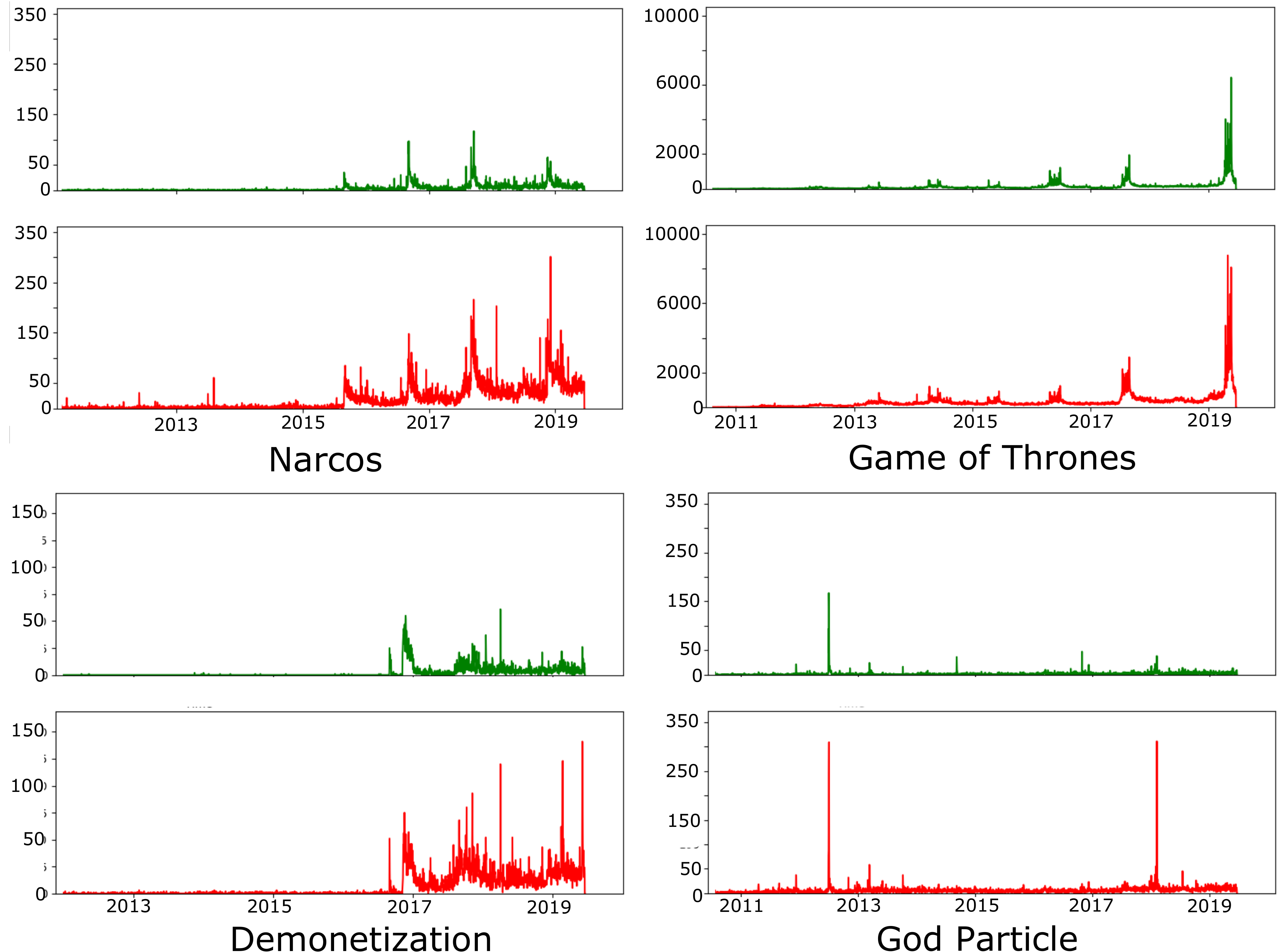}
\caption{Plots showing Reddit submissions (in green) and comments (in red) posted per hour related to four different keywords, changing over time. Y-axis in all eight plots represent absolute counts. Each keyword corresponds to a different exogenous event. We can observe spikes in the Reddit activity coinciding with the event timings.}
\label{fig:event_submission_comment}
\par\medskip
\centering
\includegraphics[width=\columnwidth]{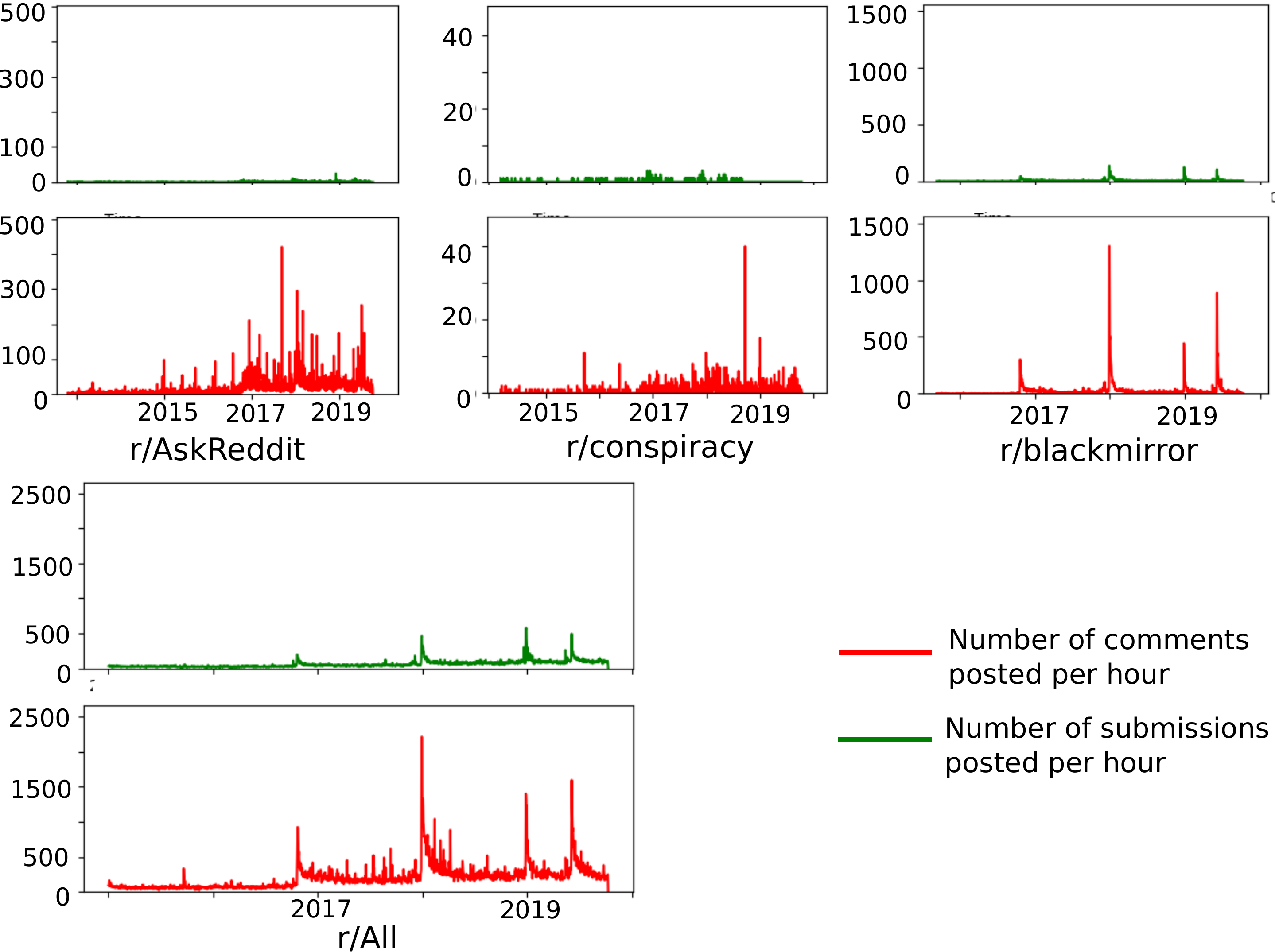}
\caption{Hourly submission and comment count for the keyword {\em Black Mirror} (a popular Web-series). We show the counts in three different communities (subreddits), and across whole of Reddit. We can clearly observe the differences in user reaction towards the same event in different communities.}
\label{fig:subreddit_sub_comment}
\end{figure}

\paragraph{\bfseries Exogenous effects.} A second critical modeling issue is {\em exogenous} influence. An influential event (recurring or sporadic) in the real world determines what topic will be the ``talk-of-the-town'' on the social media. In Figure~\ref{fig:event_submission_comment}, we show how the activities in Reddit change over time for four different topics, each corresponding to different events. {\em Narcos} and {\em Game of Thrones} are popular television series, releasing each season (collection of episodes) periodically. For both of these recurring events, there are sudden spikes in the number of submissions and comments. The last spikes in activities correspond to the final seasons where we can observe a huge gap in the number of submissions and comments, corresponding to more extensive discussions (high comment per submission ratio). The discovery of Higgs Boson (2012) and its decay (2018) corresponds to the two abrupt spikes in the activity plots of {\em God Particle}. {\em Demonetization} was an event of national importance in India in November 2016. Owing to its long-standing effects and several events triggered by it, we can observe multiple spikes even after the initial one.  It was intensely discussed during the parliamentary elections of India in 2019, as reflected in the large spikes with a high comment-to-submission ratio in the first quarter of 2019. {\em Demonetization} is a perfect example of the superposition of multiple exogenous events, financial and political.  From these examples, it is quite evident that a model that takes these exogenous events into consideration during training may result in better user engagement predictions.

\paragraph{\bfseries Endogenous (within-community) influence.} While external events play a crucial role in determining ``what people will talk about'', the degree to which such `chatter' will evolve, and the time it will take to decay, are strongly dependent on the internal or {\em endogenous} states of a community. Figure~\ref{fig:subreddit_sub_comment} shows activity levels related to {\em Black Mirror} in different subreddits.  Whereas the whole of Reddit ({\em r/all}) and the dedicated subreddit {\em r/blackmirror} follow the recurring pattern of event arrival (release of seasons), this is not the case for {\em r/AskReddit} and {\em r/conspiracy}. Interestingly, the subreddit {\em r/blackmirror} was created at the end of 2016. Flocking of dedicated followers to this subreddit and their intense engagement played a pivotal role in raising the topic's popularity in other subreddits and in Reddit overall (sharp spikes in Figure~\ref{fig:subreddit_sub_comment} after 2017 compared to the previous releases in 2014--15).

\paragraph{\bfseries \framework.}
We present \framework, a system for chatter prediction that handles the combined challenges of unknown influence network structure and exogenous influence.  \framework{} observes news and discussion streams for a limited time window up to a time \textbf{horizon}, after which it predicts \textbf{chatter}: the intensity of subsequent discussion up to another specified time.

(For concreteness, throughout this paper, we will use \textbf{news} articles as the prototypical exogenous influence.  We will use \textbf{submissions}, \textbf{comments}, and \textbf{discussion} as prototypical social network activity.  Note that these are broad model concepts that may be embodied differently in other chatter prediction applications.)

\framework{} achieves our goal using a network architecture inspired by a \textbf{unified chatter model}.   In this model, each user follows a two step process of {\em read and react}. Upon the arrival of a discussion item, a user reads its content (or views images or video). Then, depending on his/her cognitive state and the content features (topic, complexity, opinion), s/he decides whether or not to react and contribute to the discussion.  Any contribution, in turn, affects the state of other users.  This process, ``read and react'', aggregated over uesrs, can be conceptualized as an evolving mapping of content to its virality, conditioned on the dynamics of the exogenous and the endogenous states.

Using two months of Reddit discussions on 43 different subreddits,
amounting to nine million submissions and comments,
along with 3.9 million time-aligned news articles,
we show that \framework{} makes more accurate chatter predictions compared to
recent competitive approaches based on Hawkes Processes \citep{kobayashi2016tideh},
cascades \citep{cheng2014cascadepred}, and others.
Drilling down into the subreddits and contrastic their dynamics give additional insights.

\paragraph{\bfseries Summary of contributions} Our contributions are four-fold: 
\begin{itemize}[leftmargin=*]
\item Formal specification of a new chatter prediction problem in settings where social network knowledge is absent and exogenous influence is present.
\item Design and implementation of \framework, a new chatter prediction system that targets the above setting.
\item Extensive experimental comparison against prior chatter prediction methods, demonstrating the superiority of \framework.
\item New chatter prediction data set and accompanying code\footnote{Code and Sample Data at \protect\url{https://github.com/LCS2-IIITD/ChatterNet}}.
\end{itemize}

\begin{figure*}
\includegraphics[width=\textwidth]{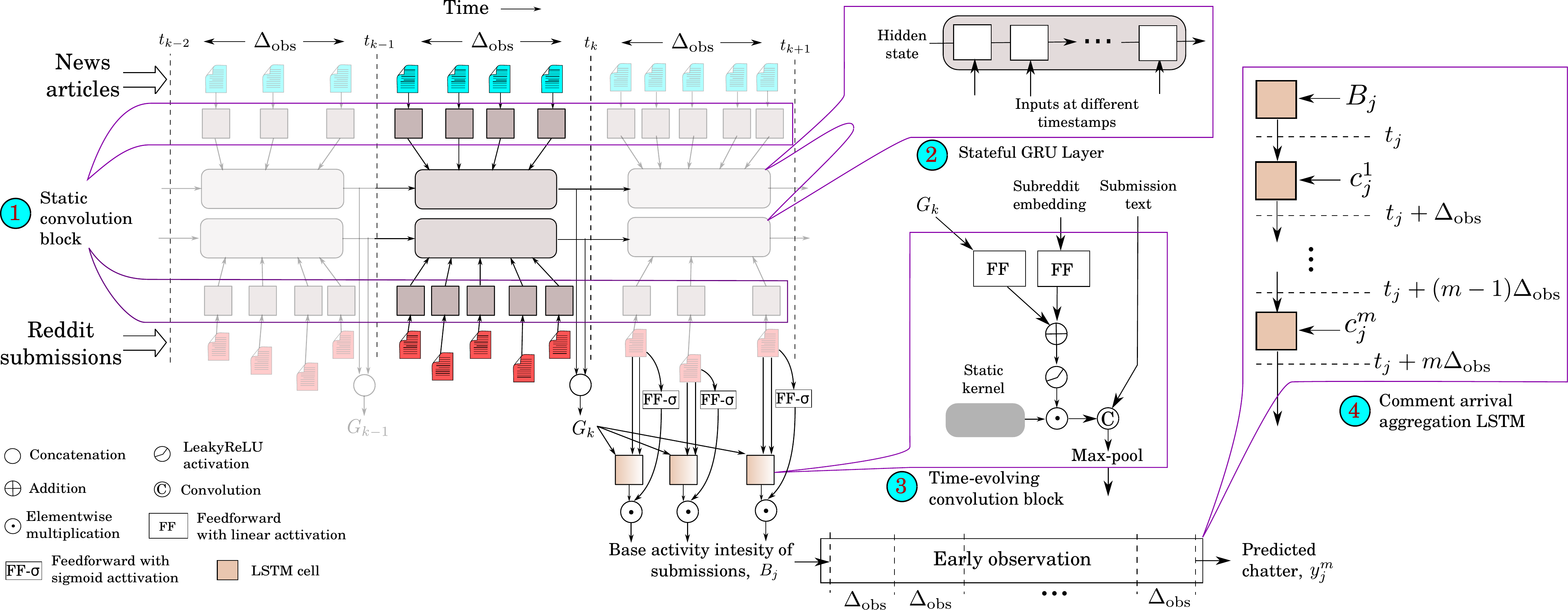}
\caption{Steps of combining exo- and endogenous influence to predict future chatter submissions over time. Indices $j$ and $k$ correspond to submissions and intervals, respectively. (i)~Features of news articles (exogenous) and submissions (endogenous), published within interval $[t_{k-1}, t_{k}]$ of length $\Delta_\text{obs}$, are extracted using a static convolution block.  (ii) Two GRUs with separate parameters aggregate each set of feature maps over the interval, and their final hidden states are concatenated to $G_k$, called the {\em time-evolving influence state}. (iii)~Shared instances of a time-evolving convolution block extract features of submissions made in $[t_k, t_{k+1}]$, controlled by $G_k$ and the subreddit embedding of the submissions. Every submission also includes average commenting rate in the corresponding subreddit, which is passed through a feedforward layer with sigmoid activation and multiplied with the outputs of time-evolving convolutions to compute base activity intensity $B_j$ of each submission. Finally, (iv)~the number of comments during each of $m$ next time intervals is aggregated with $B_j$ to predict future chatter~$y^m_j$.}
\label{fig:whole_model}
\end{figure*}


\section{Design of \framework}
\label{sec:method}

Guided by the unified chatter model discussed in Section~\ref{sec:intro},
we describe in this section the complete working of \framework.
We set up notation in Section~\ref{subsec:Method:Prelim}, and discuss some basic approaches that cannot capture all the signals available in our setting.
We describe how \framework{} combines these signals from news and discussion, in Section~\ref{subsec:influence_aggregate}.
Then, in Section~\ref{subsec:evolving_conv} we motivate why events arriving after the horizon need time-evolving network components, and present a suitable network for processing post-horizon events.  We tie these pieces together with a training loss in Section~\ref{subsec:final_prediction}.  Figure~\ref{fig:whole_model} shows a sketch of \framework.

\subsection{Preliminaries and Notation}
\label{subsec:Method:Prelim}

Time is quantized into observation intervals of length $\Delta_\text{obs}$ (e.g., $[t_{k-1}, t_k]$ in Figure~\ref{fig:whole_model}).  Intervals are indexed with~$k$.
We denote as $\hat{N}:=\langle (n_i, t_i)|\forall i\in \mathbb{Z}^+\rangle$
the stream of news articles $n_i$ with publication timestamps~$t_i$. 
Let $\hat{S}:=\langle (s_j, t_j)|\forall j\in \mathbb{Z}^+\rangle$
be a stream of submission items $s_j$ posted at time~$t_j$. Every news item $n_i$ consists of a text digest with headline and body, while every submission item $s_j$ is a triplet $(s^T_j, s^V_j, s^R_j)$, where $s^T_j$ is the text digest of $s_j$, $s^V_j$ is the subreddit to which $s_j$ was posted, and $s^R_j$ is the average commenting activity (number of comments) in the subreddit within the previous interval.
For any submission $s_j$ posted at timestamp~$t_j$ with $t_k<t_j\leq t_{k+1}$, we define an {\bf observation window} $[t_j, t_j+m \Delta_\text{obs} ]$ and a {\bf prediction window} $[t_j+m\Delta_\text{obs}, t_j+\Delta_\text{pred}]$, where $m\in \mathbb{N}$ is an application-driven hyperparameter. Note that, in this setting it is important to differentiate the roles of submissions posted before and after $t_k$, which is the boundary up to which we have the most recent exogenous-endogenous signals defined. So every submission up to $t_k$ contributes to this endogenous signal. The submission posted within $t_k<t_j\le t_{k+1}$ lies between the two arrivals of influence signal, one in the past ($t_k$) and one in the future ($t_{k+1}$). So, every such submission might be under the influence signal at~$t_k$.

Our goal is to predict \textbf{chatter} pertaining to $s_j$, defined as $y_j = \ln (1+C_j)$, where $C_j$ is the total number of comments made about $s_j$ within the prediction window, after observing commenting activity about $s_j$ within the observation window.  (Additive error in log-count prediction amounts to count prediction within a multiplicative factor.  $C_j$ depends on $\Delta_\text{pred}$ but we elide that for simpler notation.)
As early predictions are most beneficial, we specify two settings:
\begin{description}[leftmargin=*]
\item[Zero-shot:] Empty observation window, with $m=0$.
\item[Minimal early observation:] Here $m>0$, but $m\Delta_\text{obs} \ll \Delta_\text{pred}$.
\end{description}
Popular time-series models cannot perform the above tasks well, for several reasons.  First, any static mapping from the textual features of the submissions to their corresponding future chatter fails to incorporate the dynamic exogenous and endogenous influences that govern chatter. Second, generative models need a substantial degree of early observation which is not available in our setting. Third, the high arrival rates demand fast response.  Therefore, when predicting the future chatter of a submission, chatter under its predecessors in the stream remains mostly unobservable.  This precludes autoregressive \citep{kobayashi2016tideh} approaches.  Finally, the lack of knowledge of user-user ties inhibits the employment of information diffusion models~\citep{cheng2014cascadepred}.

The cumulative aggregate influence of exogenous and endogenous signals from news and submission streams during $[t_{k-1}, t_k]$ will be endowed a deep representation $G_k$, as defined in Section~\ref{subsec:influence_aggregate}.  $G_k$ will help map submission texts posted within the next interval to a base chatter intensity, which will be aggregated with the activity within the observation window to predict the final chatter.

In what follows, every weight and bias matrix (denoted as $W$ and $Q$, respectively with various subscripts) belongs to the trainable model parameter set. Every text segment (news and submission) is mapped to a sequence of low dimensional representations using a shared word embedding layer. Moreover, the subreddit information $s^V_j$ corresponding to each submission $s_j$ is mapped to a \textbf{subreddit embedding vector} $\mathbf{U}_j$ using a shared embedding layer. All these embeddings are part of the trainable parameters of \framework.

\subsection{Cumulative Influence Aggregation}
\label{subsec:influence_aggregate}

We define two functions that compute exogenous and endogenous influences:
\begin{align}
\text{Exogenous:} & \quad
\mathcal{F}_\text{X}(N|t_i\leq t \ \forall (n_i, t_i)\in N\subseteq \hat{N})\\
\text{Endogenous:} & \quad
\mathcal{F}_\text{E}(S|t_j\leq t \ \forall (s_j, t_j)\in S\subseteq \hat{S})
\end{align}
We model each of these functions $\mathcal{F}_\text{X}$ and $\mathcal{F}_\text{E}$ in two steps.

First, we map each text digest to its influence feature map using a shared convolution block (refer to component ($1$) in Figure~\ref{fig:whole_model}). Given an input sequence of word vectors of a news (submission text) as $n_i$ ($s^T_j$), we apply successive 1-dimensional convolution and max-pooling operations to produce a feature map $X^n_i$ ($X^s_j$):
\begin{equation}
\label{eq:static_conv1d}
    \begin{split}
    C^n_i &= \text{ReLU}\bigl(\text{Conv1D}(n_i|W_\text{static}) \bigr)\\
    X^n_i &= \text{MaxPool}(C^n_i)
    \end{split}
\end{equation}
where $W_\text{static}$ denotes filer kernels and $C^n_i$ ($C^s_j$) is an intermediate representation. We use parallel branches of convolution and pooling operations with different kernel sizes (1, 3, and 5) to capture textual features expressed by contexts of different sizes. The outputs from each branch are then concatenated to produce the final feature representation $f^n_i$ ($f^s_j$) corresponding to $n_i$ ($s_j$) (detailed organization explained in the Appendix, Figure~\ref{fig:text_cnn}). The submission feature maps are concatenated with subreddit vector $\mathbf{U}_j$ to differentiate the influences of submissions from different subreddits, as discussed in Section~\ref{sec:intro}. The final feature map corresponding to $s_j$ is then $f^{sv}_j$.  We denote these stages as {\em static} because $W_\text{static}$ remains temporally invariant.

\paragraph{\bfseries ConvNet vs.\ LSTM/BERT}
We chose a simple convolutional architecture \citep{kalchbrenner2014convnetsentence, kim2014convnetsentence} over a recurrent one for two reasons --- i)~convolution reduces the size of parameter space, which is essential in handling large data streams, and ii)~our task requires efficient understanding of topic-specific keywords to constitute the influence, as opposed to the complex linguistic structures with long-term dependencies; this makes recurrent architectures an overkill. Our experiments with BERT~\citep{DBLP:conf/naacl/DevlinCLT19} instead of convolution to produce text representations (expectedly) gave no significant gain.
\par\smallskip

Next, using the convoluted feature maps $f^n_i$ ($f^{sv}_j$), we compute discrete approximations of the functions $\mathcal{F}_\text{X}$ ($\mathcal{F}_\text{E}$) at the end of every interval $[t_{k-1}, t_k]$ as $G^n_k$ ($G^s_k$), as follows (see component (4) in Figure~\ref{fig:whole_model}):
\begin{equation}
    G^n_k = \text{GRU}\Bigl(h^n_{k-1}, \langle f^n_i|t_{k-1}<t_i\leq t_{k}\rangle | W^n_G\Bigr)
\end{equation}
where $\text{GRU}$ is a Gated Recurrent Unit, $t_i$ ($t_j$) corresponds to the timestamp associated with $n_i$ ($s_j$), $h^n_{k-1}$ is the hidden state of the GRU from the previous interval, and $W^n_G$ ($W^s_G$) is the parameter set. Stateful propagation of the hidden state ensures the modeling of short-term as well as long-term influence signals.
Finally, the cumulative influence $G_k$ is computed as the concatenation of $G^n_k$ and~$G^s_k$.

We deploy two different GRUs to aggregate the exogenous and endogenous influences over the intervals given the different arrival patterns of news articles over web and submissions over Reddit (news come in sparse bursts, while submissions mostly come in very high rate).

Again, our choice of GRU as the recurrent information processing layer for this task is motivated by our experiments confirming LSTMs to be slower with no performance gain compared to GRUs.

\subsection{Time-evolving Convolution (TEC)}
\label{subsec:evolving_conv}

Following the intuitive motivation of the unified chatter model, we may now seek to map any submission $s_j$ posted in the interval $[t_k, t_{k+1}]$ to its potential to invoke future chatter, controlled by the cumulative influence from the previous interval. Formally, this mapping can be defined as,
\begin{equation}
    \mathcal{B} = \mathcal{F}_G(s_j|G_k, \ t_k<t_j\leq t_{k+1})
\end{equation}

We again resort to 1-dimensional convolution to learn feature maps from $s_j$, but this time, the filter kernel $W_\text{TEC}$ being a function of $G_k$ and the subreddit vector $\mathbf{U}_j$ corresponding to $s_j$:
\begin{equation}
    W_\text{TEC} = W_S\odot\gamma (W_G\cdot G_k + W_V\cdot s^V_j)
\end{equation}
where $\gamma(x)=x$ if $x\ge 0$, and $\alpha x$ of $x < 0$ (standard LeakyReLU activation with parameter $\alpha$, experimentally set to $0.2$). $W_G$ and $W_V$ (corresponding to the two feed-forward layers inside component (3) in Figure~\ref{fig:whole_model}) control the contributions of the cumulative influence and the subreddit, respectively. This $\gamma (W_G\cdot G_k + W_V\cdot s^V_j)$ component `calibrates' the static kernel $W_S$ with the element-wise multiplication according to the influence. As $G_k$ evolves over time, so does $W_\text{TEC}$.

Equipped with this influence-controlled kernel, the time-evolving  convolution and max-pooling on $s_j$ can be defined similar to Eq.~\ref{eq:static_conv1d}. Again, we apply parallel branches of successive convolution with different filter sizes and max-pooling, concatenation, and another series of convolutions (complete organization shown in Appendix, Figure~\ref{fig:ada_cnn}) to finally map $s_j$ to a non-negative real value $\Tilde{B}_j$, the {\bf potential chatter intensity} of $s_j$ independent of the subreddit where $s_j$ is posted.

\subsection{Final Prediction}
\label{subsec:final_prediction}

Chatter levels in different subreddits vary with the number of active users at any time. The average commenting activity $s^R_j$ of the subreddit corresponding to $s_j$ enables us to compute the relative activity signal $r_j \in \mathbb{R}$ such that,
\begin{equation}
    r_j = \sigma(W_R\cdot s^R_j + Q_R)
\end{equation}
where $\sigma(x)= (1+e^{-x})^{-1}$ (shown as ``FF-$\sigma$'' blocks in Figure~\ref{fig:whole_model}). We compute the {\bf base chatter intensity} corresponding to $s_j$ as $B_j = r_j\Tilde{B}_j$ such that $0<r_j<1$ plays the role of a scaling factor to calibrate $B_j$ according to the activity level of the subreddit.


Having computed $B_j$, the chatter intensity invoked by $s_j$,
under the influence of past history of news and submission arrival and calibrated by the subreddit information, we next observe the commenting activity under $s_j$ within the observation window (see component (4) in Figure~\ref{fig:whole_model}). We employ a binning over time intervals to transform the comment arrivals within the observation window $[t_j, t_j+m\Delta_\text{obs}]$ into a sequence $\langle c^1_j, c^2_j, \cdots, c^m_j\rangle$ where each $c^l_j$ is the total number of comments arrived within $[t_j+(l-1)\Delta_\text{obs}, t_j+l\Delta_\text{obs}]$. This sequence serves as a coarse approximation of the rate of comment arrivals over the observation window. We use a single LSTM layer to aggregate this sequence and predict the final chatter $y^m_j$ (superscript corresponds to the length of the observation window). In the zero-shot setting (i.e., $m=0$) this LSTM is not used and we predict chatter $y^0_j = B_j$.

Details of \framework\ parameters are given in Appendix~\ref{appendix:parameters}.
\subsection{Cost/Loss Functions}
\label{subsec:cost_function}

In a realistic setting, there are far too many discussions invoking near-zero chatter along with very small number of those which go viral heavily. As we take both of these types without any filtering (opposed to excluding less viral ones in the cascade prediction tasks like \citep{kobayashi2016tideh} or \cite{zhao2015seismic}), the cost function needs to handle skewed ground truth values. To deal with this, we train \framework\ by minimizing the mean absolute relative error given by $\sum_j \frac{\lvert y_j - y^m_j \rvert}{y_j+\epsilon}$, where $y_j$ is the chatter ground truth, $y^m_j$ is the predicted chatter, and $\epsilon$ is a small positive real number to avoid division by zero (as implemented in Keras/Tensorflow).

\begin{table}[h]
\caption{Feature set for the baseline {\em CasPred}, where $T$:= set of unique terms in corpus, $tf_t$:= term frequency of term $t\in T$ in the submission, $|w|$:= number of words in the submission, $|cw|$:= number of words in the submission with more than 6 letters, $|s|$:= number of sentences in the submission, $k$:= observable discussion size, $t_i$:= time when $i$-th comment was put, $t_0$:= time of submission. * signifies a feature not in the original paper but added by us.}
\small
\begin{tabular}{|l|l|}
\hline
{\bf Features} & {\bf Expression} \\ \hline
Bag-of-words * &  Unigram features with tf-idf\\ \hline
Complexity * & \makecell[l]{Degree of unique tokens used,\\ $
         p=\frac{1}{|T|}\sum_{t\in T}tf_t(\log|T|-\log(tf_t))$}\\ \hline
LIX Score * & \makecell[l]{Readability score of the submission,\\ $r = \frac{|w|}{|s|}+100\times\frac{|cw|}{|w|}$} \\ \hline
Polarity & \makecell[l]{Sum of sentiment intensity scores \\ of the unique terms of the submission,\\ computed using SenticNet~\citep{cambria2018senticnet}} \\ \hline
Referral count * & Number of URLs in the submission \\ \hline
Size * & \makecell[l]{Number of words and \\ sentences in the submission} \\ \hline
Subreddit * & In which subreddit the submission is posted \\ \hline
Commenting time &  \makecell[l]{Time elapsed between the \\ $i$-th comment and the submission}\\ \hline
\begin{tabular}[c]{@{}l@{}}Average time difference\\ in first $k/2$ comments\end{tabular} & $\frac{1}{k/2-1}\sum_{i=1}^{k/2-1}(t_i-t_{i-1})$ \\ \hline
\begin{tabular}[c]{@{}l@{}}Average time difference\\ in last $k/2$ comments\end{tabular} &  $\frac{1}{k/2-1}\sum_{i=k/2}^{k}(t_i-t_0)$\\ \hline
\end{tabular}
\label{tab:cascade_features}
\end{table}

\section{Experiments}
\label{sec:experiment}

\subsection{Dataset}
\label{subsec:data}
We collected the discussion data from Pushshift.io\footnote{\protect\url{https://files.pushshift.io/reddit/}}, a publicly available dump of Reddit data, stored in monthly order. We used the discussion data of October and November 2018, from 43 different communities (Subreddits). The October data was used for training and development, and the November data was used for testing. In total, we have a collection of 751,866 submissions with 2,604,839 comments in the training data, and 1,334,341 submissions with 4,264,177 comments in the test data.

To fetch the news articles published online, we relied on the news-please crawler \citep{Hamborg2017}, which extracts news articles from the Common Crawl archives\footnote{\protect\url{https://commoncrawl.org/}}. We crawled the news articles published in the same timeline as of the Reddit discussions. We got 1,851,022 articles from $4757$ different news sources for the month of October, and 2,010,985 articles from $5054$ sources for November. It should be noted that this covers all the news articles published in English in this period, as we chose not to use non-English data. Details of preprocessings are given in Appendix~\ref{appendix:preprocess}.

\subsection{Training and Evaluation Protocols}
\label{subsec:param_select}

For training the model, we use the October data divided into the train-validate split. The news and discussion streams in the period from \formatdate{3}{10}{2018} \formattime{0}{0}{0} GMT to \formatdate{22}{10}{2018} \formattime{23}{59}{59} GMT are used for training \framework, while the validation is done using the data from \formatdate{23}{10}{2018} \formattime{0}{0}{0} GMT to \formatdate{30}{10}{2018} \formattime{23}{59}{59} GMT. We set $\Delta_\text{obs}$ and $\Delta_\text{pred}$ to $60$ seconds and $30$ days, respectively. We train multiple variations of \framework\ with different observation windows: $15$, $30$, $45$, and $60$ minutes ($m=15, 30, 45, 60$).  Additional training detail are given in the Appendix~\ref{appendix:training}.

\subsection{Baseline Models}
\label{subsec:baseline}

Due to the novelty of the problem setting and the absence of social network information, comparing \framework\ with the state-of-the-art is not straightforward. We engage four external baselines for retweet cascade prediction and Reddit user engagement prediction, tailored to our setting. We also implement multiple variants of \framework\ for extensive ablation analysis of the different signals.

\subsubsection{\bfseries TiDeH}
To adapt Time Dependent Hawkes Process \cite{kobayashi2016tideh} as a baseline in the absence of any knowledge of the underlying user network, we set the follower count of each Reddit poster/commenter as~1. In addition, we set the minimum thread size (minimum number of comments) to~10.

\subsubsection{\bfseries CasPred}
\label{subsubsec:cascade}
The \mbox{(re-)sharing} cascade prediction approach of \citet{cheng2014cascadepred} provides an interesting baseline for \framework\ by allowing us to test not only the temporal but also the textual features of our dataset. 
Due to limitations of Reddit metadata we can make use of only a subset of the features they used. We also include some additional content features fitting to discussions in Reddit \citep{DBLP:conf/icdm/Dutta0019} (complete feature set in Table \ref{tab:cascade_features}). We implement CasPred-org (original features) and CasPred-full (augmented with additional features) with observable cascade size $k=10$. 

\subsubsection{\bfseries RGNet}
Our third external baseline is an adaptation of the Relativistic Gravitational Network \citep{DBLP:conf/icdm/Dutta0019}, primarily designed to predict user engagement behavior over Reddit.

\subsubsection{\bfseries DeepCas}
DeepCas~\citep{DBLP:conf/www/LiMGM17} makes use of the global weighted topology of inter-user ties. Since, Reddit does not posses any explicit user-user mapping, we consider an edge between the posters and the commenters of a post (to generate the global network). Also, each post(with its set of poster and commenters) is treated as a cascade. 
\subsubsection{\bfseries Ablation variants of \framework}
\label{subsubsec:ablation}
To observe the contribution of different components of \framework, we implement the following ablated variants:

    $\bullet$ {\bf \framework-N} which uses only the news-side influence signal;
    
    $\bullet$ {\bf \framework-S} which uses only the submission-side influence signal;
    
    $\bullet$ {\bf \framework-Static} which does not use any influence signal; for this, the time-evolving convolution block is replaced by a static convolution block;
    
    $\bullet$ {\bf LSTM-CC} which uses only the LSTM layer aggregating the observed comment arrivals (Section~\ref{subsec:final_prediction}).
LSTM-CC allows implementation for only the minimal early observation setting; zero-shot is not supported.  Other variants are implemented for both of the task settings.

\begin{table}
\caption{Evaluation of \framework\ and the external baselines (over complete test data). $\tau$ and $\rho$ correspond to Kendall's $\tau$ and Spearman's $\rho$, respectively, whereas Step-wise $\tau$ corresponds to Kendall $\tau$ on the sampled ground truth. \framework+ and \framework++ correspond to the zero-shot and minimal early observation of $1$ hour.}
\small
\begin{tabular}{|l|c|c|c|c|}
\hline

{\bf Model} & {\bf MAPE} & $\mathbf{\tau}$ & $\mathbf{\rho}$ & {\bf Step-wise} $\mathbf{\tau}$ \\ \hline
\framework+ & 33.142 & 0.4042 & 0.4601 & 0.8781  \\ \hline
\framework++ & {\bf 25.893} & {\bf 0.4439} & {\bf 0.5050} & {\bf 0.8980} \\ \hline
TiDeH (1hr.) & 35.178 & 0.0715 & 0.1140 & 0.5622 \\ \hline
CasPred-full & - & - & - & 0.4741 \\ \hline
CasPred-org & - & - & - & 0.3515 \\ \hline
RGNet & 148.34 & 0.1871 & 0.2273 &  0.5305\\ \hline
DeepCas & 163.6 & 0.2362 & 0.3309 &  0.2636\\ \hline
\end{tabular}
\label{tab:result_overall_agg}
\end{table}

\begin{table}
\caption{Performances of different ablation variants of \framework\ (see Section~\ref{subsubsec:ablation}). Except for LSTM-CC, all the variations are tested for zero-shot (ZS) and early observation of 1 hour (OBS). MAPE and $\tau$ are as mentioned in Table~\ref{tab:result_overall_agg}.}
\small
    \centering
    \begin{tabular}{|l|c|c|c|}
        \hline
        {\bf Model} & {\bf Setting} & {\bf MAPE} & $\tau$\\\hline
        \framework & \makecell{ZS\\OBS} & \makecell{33.142\\25.893} & \makecell{0.4042\\0.4439} \\\hline
        \framework-N & \makecell{ZS\\OBS} & \makecell{38.498\\29.023} & \makecell{0.3267\\0.4134} \\\hline
        \framework-S & \makecell{ZS\\OBS} & \makecell{37.007\\28.890} & \makecell{0.3511\\0.4201} \\\hline
        \framework-Static & \makecell{ZS\\OBS} & \makecell{195.344\\149.558} & \makecell{0.1488\\0.3580} \\\hline
        LSTM-CC & OBS & 152.313 & 0.3580 \\\hline
    \end{tabular}
    \label{tab:ablation}
\end{table}

\begin{table*}
\caption{(Ablation study) Subreddit-wise MAPE for \framework, \framework-N, and \framework-S all using early observation window of size 1 hour. Subreddits with * are abbreviated names given by:  AR$\rightarrow$ AskReddit, TD$\rightarrow$ The\_Donald, WITT$\rightarrow$ whatisthisthing, RL$\rightarrow$ RocketLegue, TNF$\rightarrow$  TheNewsFeed, FF$\rightarrow$ Fantasy\_Football, NSQ$\rightarrow$ NoStupidQuestions, AS$\rightarrow$ askscience, WSB$\rightarrow$ wallstreetbets, GO$\rightarrow$ GlobalOffensive, PF$\rightarrow$ personalfinance, UPO$\rightarrow$ unpopularopinion, EI$\rightarrow$ EcoInternet, AN$\rightarrow$ AutoNewspaper, TTF$\rightarrow$ TheTwitterFeed, NBB$\rightarrow$ newsbotbot, NBTMT$\rightarrow$ newsbotTMT, NBMARKET$\rightarrow$ newsbotMARKET, BN24$\rightarrow$ BreakingNews24hr, PH$\rightarrow$ PoliticalHumor, BCAll$\rightarrow$ BitcoinAll. Results for some of the subreddits with significant changes due to ablation of exogenous/endogenous signals are highlighted in bold.}
\small
\begin{tabular}{|l|c c c c c c c c c c c|}
\hline
{\bf Model} & {\bf AR*} & {\bf TD*} & {\bf gaming} & {\bf politics} & {\bf technology} & {\bf Music} & {\bf techsupport} & {\bf WITT*} & {\bf news} & {\bf movies} & {\bf RL*} \\ \hline
\framework &   {\bf 26.031}  & 31.854     &     24.462  &   {\bf 23.389}    &    25.686   &  21.249   &    25.13 &  26.002     & {\bf 29.102}     &     25.076  &    24.221   \\ 
\framework-S & {\bf 28.411}   & 34.712   &   31.011     &   {\bf 38.105}    & 36.122      &    28.310   & 36.54    &   27.151  & {\bf 35.671}     &   28.510     &      27.907      \\ 
\framework-N  &  {\bf 30.008} & 35.003   &  28.949      &   {\bf 25.610}    &   30.991    &  27.569   &  29.171   &  27.711    & {\bf 29.342}   & 26.134  &      27.886      \\ 
     \hline
{\bf Model} & {\bf Tinder} & {\bf TNF*} & {\bf anime} & {\bf india} & {\bf Jokes} & {\bf soccer} & {\bf FF*} & {\bf NSQ*} & {\bf nfl} & {\bf AS} & {\bf WSB*} \\ \hline
\framework &   23.861        &       33.420 &   26.004  & {\bf 27.009}  &  25.562    & {\bf 24.753}     & 27.419     &    25.510    &   28.911   & 23.70  &   23.251      \\ 
\framework-S &     25.945      &     36.138 &   28.040  & {\bf 39.202}  &  32.787 & {\bf 33.402}  &  29.958   &    29.875        &   35.007   &   26.419      &      24.36      \\ 
\framework-N &    26.020       &     36.287 &   29.011  & {\bf 32.784}  &  34.095   &  {\bf 30.92}   &  29.011   &   31.592     &  32.019    &  24.499         &    24.212        \\ 
\hline
{\bf Model} & {\bf InNews} & {\bf GO*} & {\bf teenagers} & {\bf POLITIC} & {\bf brasil} & {\bf NBA2k} & {\bf bussiness} & {\bf PF*} & {\bf nba} & {\bf worldnews} & {\bf UPO*} \\ \hline
\framework &  29.534   & 24.993 &    24.771    &  27.364  &   24.77    & 25.333 & 26.001  &    23.122    &   25.251   &   24.924    &       28.037     \\ 
\framework-S &  32.119 & 27.604 &    26.759    &  29.990 &  26.904     &  30.424  &  31.213 &   31.797  & 28.751  &     28.117    &       30.301     \\ 
\framework-N &   30.54  &  27.601  &  27.002  &  32.013    &  27.591     & 28.701    &  28.10   &    32.107    & 26.29     &        25.023         &       29.213     \\  \hline
{\bf Model} & {\bf EI*} & {\bf AN*} & {\bf NBB*} & {\bf FIFA} & {\bf BN24*} & {\bf BCAll*} & {\bf NBTMT*} & {\bf TTF*} & {\bf PH*} & {\bf NBMARKET*} & - \\ \hline
\framework &   27.001   &  27.159 &   24.146  &  0.754  & 24.113 &  28.011 &  26.112   &   25.301   &  24.35    &         23.109        &     -       \\ 
\framework-S &  28.994  &  32.571 &   32.386  &  29.778 & 29.292 &  27.003 &  29.203   &   26.906   &  24.997   &        27.904         &        -    \\ 
\framework-N &   28.923 &  28.529 &   25.183  &  29.022 & 28.124 &  27.091 &  28.114   &   28.476   &  26.870    &         27.878        &      -      \\ \hline
\end{tabular}
\label{tab:subreddit_aggregate}
\end{table*}

\section{Evaluation}
\label{sec:evaluation}

We explore multiple evaluation strategies to see how \framework\ responds to different challenges of chatter prediction. We use three different evaluation metrics: a)~Mean Absolute Percentage Error (MAPE), b)~Kendall rank correlation coefficient  (Kendall's $\tau$), and c)~Spearman's $\rho$. As the CasPred model does not predict the exact size of the discussion but gives a binary decision of whether a given submission will reach at least size $l\times k$, $l\in \mathbb{Z}^{+}$ after observing a growth of size $k$, we can not evaluate this with the mentioned three metrics directly. Instead, we map the ground-truth to these $l\times k$ values such that the label of a discussion with size $d$ would be $\left \lfloor{\frac{x}{k}}\right \rfloor $. Then we compute Kendall's $\tau$ over this values (hereafter called as step-wise~$\tau$). We use the same binning (with $k=10$) for rest of the models to evaluate the step-wise~$\tau$.

\subsection{Overall Performance}

In Table~\ref{tab:result_overall_agg}, we show the evaluation results for \framework\ in zero-shot and early observation settings along with all the external baselines. While \framework\ exploiting comment arrival within the early observation window outperforms rest of the  models by a large margin, it also performs better than the external baselines in the zero-shot setting.

An interesting pattern can be observed with TiDeH, RGNet and DeepCas. While TiDeH produces predictions comparable to \framework\ in terms of MAPE, it suffers largely in terms of rank correlation. On the other hand, both RGNet and DeepCas follow a completely opposite pattern -- better ranking of future chatter compared to predicting the actual value of chatter.

The poor performance of CasPred and RGNet can be explained in terms of the difference between their original design context and the way they are deployed in our problem setting. Almost two-third of the feature set originally used for CasPred can not be implemented here. Also, it is evident that our additional feature set actually improves the performance of CasPred, signifying the importance of these features for engagement modeling in Reddit. In case of RGNet, it is built to take into account the dynamics of user engagement over time. But in the absence of a rich feature set and social network information, it is the use of endogenous and exogenous influence which gives \framework\ such leverage compared to the baselines.

Earlier we highlighted the major challenge of predicting future chatter without delayed observation of chatter evolution. \framework\ satisfies this requirement better than other baselines, because they were all designed for much larger early observation windows.   \emph{TiDeH takes $24$ hours of observation to outperform \framework\ with $1$ hour of observation.}  Via time-sensitive combination of exogenous and endogenous signals, \framework\ achieves superior performance without network knowledge.

\subsection{Ablation of \framework\ Components}
\label{subsec:ablation_results}

We justify the complexity of \framework, and show that all its pieces are critical.
In Table~\ref{tab:ablation}, we present the performances of the various ablation models described in Section~\ref{subsubsec:ablation}. It is evident that removal of either exogenous or endogenous signals from \framework\ results in a degraded performance. However, \framework\ with only endogenous signal slightly outperforms its counterpart with only exogenous signal. This difference does not tell us whether endogenous signals are more important --- we need to study their effect for individual subreddits to comment on that.

Removing both signals degrades performance heavily, particularly in the zero-shot setting. This is expected, because \framework-Static in the zero-shot regime is simply a static convolution block mapping submission texts to their future chatter -- a regression task juxtaposed with simple text classification engine. The decrease in performance is more evident with the MAPE measure ($195.344$ and $149.558$, respectively for zero-shot and early observation). Even in zero-shot setting \framework-Static utilizes the information of average comment arrival in the subreddit to scale the future chatter accordingly and learn at least a possible ranking of submissions with respect to their future chatter. Additionally we removed this operation as well for \framework-Static; kendall $\tau$ for this further ablated model dropped to $0.02$.

Comparing the performances of \framework-Static in zero-shot (only submission features) and LSTM-CC (only comment arrival features), one can easily conclude that, when the exogenous and endogenous signals are not taken into account, comment arrival patterns are much powerful indicators of future chatter compared to submission texts. 

\subsection{Effect of Observation Window and Size of Discussion}

\begin{table}[!t]
\caption{MAPE and kendal-$\tau$ scores to predict future chatter using \framework\ and LSTM-CC, each with varying size of the observation window. {\em model\_name-$x$} signifies the model uses an observation window $x$ minutes long.}
    \centering
    \begin{tabular}{|l|c|c|}
    \hline
{\bf Model} & {\bf MAPE} & $\tau$ \\
\hline
\framework-0   & 33.142 & 0.4042 \\
\framework-15   & 31.886 & 0.4278 \\
\framework-30   & 28.12 & 0.4302 \\
\framework-45   & 26.042 & 0.4361 \\
\framework-60   & 25.893 & 0.4439 \\
\hline
LSTM-CC-15   & 196.128 & 0.0639 \\
LSTM-CC-30   & 174.667 & 0.1307 \\
LSTM-CC-45   & 167.024 & 0.2259\\
LSTM-CC-60   & 152.313 & 0.3580\\
\hline
    \end{tabular}
    \label{tab:obs-window}

\end{table}

\begin{figure}
\centering
\includegraphics[width=0.7\columnwidth]{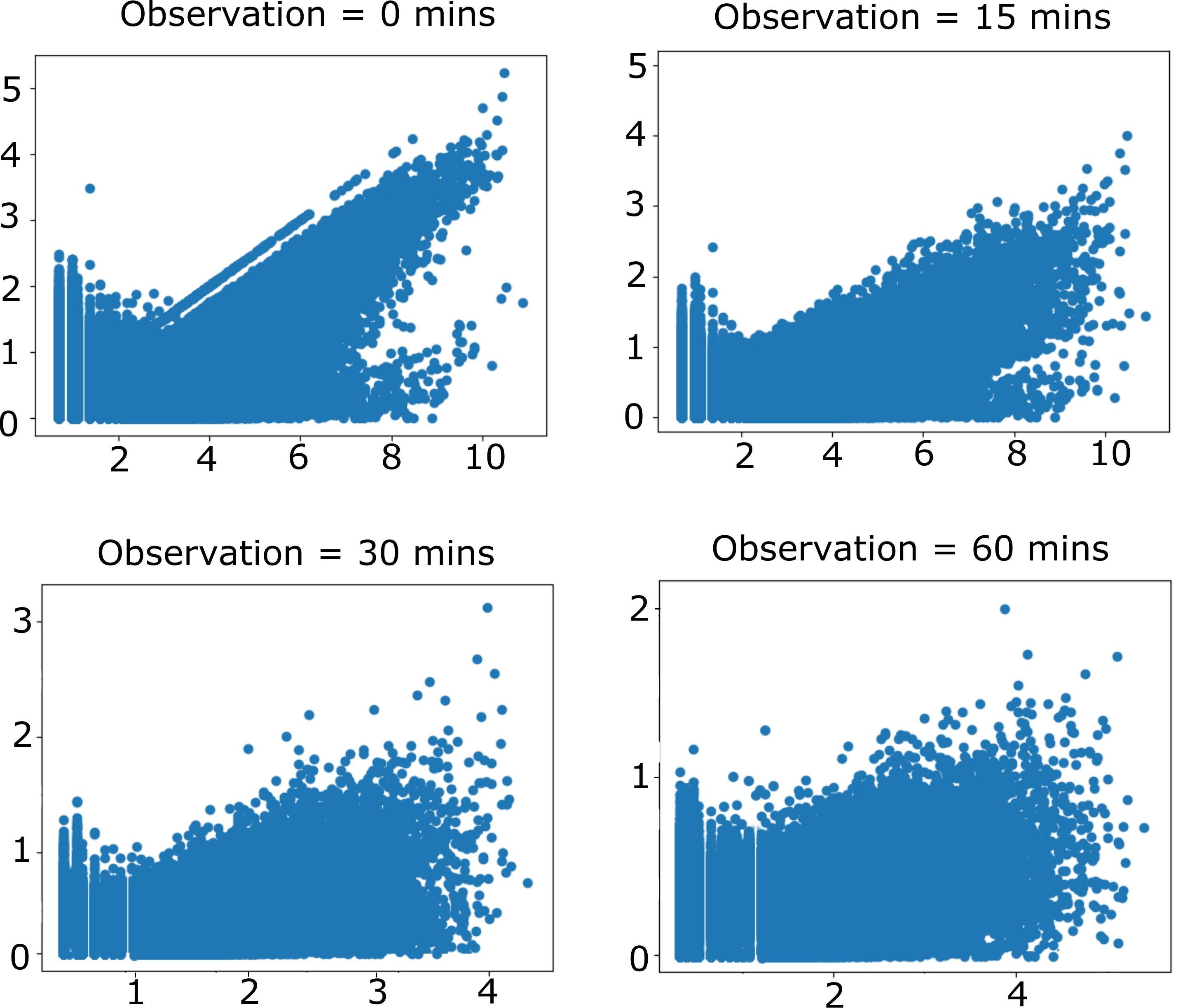}
\caption{Plots showing absolute error (y-axis) vs. ground truth value (x-axis, log of aggregate comment count to a submission) for \framework\ designed with different observation windows. With increasing observation window, the volume of discussion remaining to predict (total size $-$ observation size) decreases, corresponding to the shrinking x-axis in the plots.  As observation window increases, the error vs. gold value slope decreases, i.e., higher values of chatter become more predictable}.
\label{fig:observation_vs_size}
\vspace{-3mm}
\end{figure}

Table~\ref{tab:obs-window} shows the variation of performance for \framework\ and LSTM-CC with different sizes of observation window used to aggregate early arrivals of comments. While LSTM-CC shows a steady betterment of performance with increasing observation, \framework\ takes a quick leap from zero-shot to $15$ minute early observation and then reaches a nearly-stationary state. As shown in Figure~\ref{fig:observation_vs_size}, with longer initial observation, \framework\ tends to decrease the error rate for predicting high values of chatter.

\citet{cheng2014cascadepred} reported
increasing uncertainty in predicting larger cascades. We plot the absolute error in prediction vs. ground-truth value in Figure \ref{fig:observation_vs_size}, for different early observation windows. We measure the absolute error to predict the size gain after observation. In all four cases, absolute error varies almost linearly with size. However, with longer observation, the slope drops. With a 60 minutes long early observation, absolute error nearly grazes a zero slope line. However, these plots shows the joint effect of increasing observation and decreasing post-observation gold value. 
Table~\ref{tab:pred-window} summarizes how well \framework\ predicts future chatter with different prediction windows. Again, longer a discussion persists, harder it becomes to predict the final amount of chatter. Also, as can be expected, the zero-shot system tends to suffer more with longer prediction window.
\subsection{Subreddit-wise Analysis}
\label{subsec:subreddit_analysis}

\paragraph{\bfseries Exogenous vs. Endogenous influence} 
While Table~\ref{tab:ablation} provides useful insights about the roles played by \framework{} components, drilling down from aggregate performance into different subreddits gives additional insight.
As discussed in Section~\ref{sec:intro}, endogenous and exogenous influence manifest themselves differently over different subreddits (which is why we used subreddit embeddings $\mathbf{U}_j$ in both components: influence aggregation and time-evolving convolution).  In Table~\ref{tab:subreddit_aggregate}, we present the performances of  \framework\,  \framework-N, and \framework-S for each of the 43 subreddits.

In some subreddits (e.g., {\em r/techsupport, r/india, r/business, r/POLI\-TIC, r/InNews}, etc.), \framework\ suffers more with the ablation of exogenous signal compared to the endogenous one. Most of this subreddits are either directly news related (like {\em r/news, r/InNews, r/worldnews,} etc.), or very closely governed by what is happening in the real world, like {\em r/technology, r/business, r/nfl, r/movies,} etc.  

Some subreddits are naturally grouped, i.e., they share common topics of discussion, common set of commenting users, etc. Subreddits like {\em r/Music, r/movies and r/anime} fall into one such group. This sharing of information facilitates \framework-S to perform better compared to \framework-N as it uses the endogenous knowledge in terms of previous submissions, subreddit embeddings, and comment rates. Also there are some particular subreddits (e.g., {\em r/AskReddit, r/teenagers, r/NoStupidQuestions}) where endogenous information becomes more important than the exogenous one.

\begin{table}[]
\caption{Effect of early observation on the chatter prediction performance over different subreddits. We take \framework\ in zero-shot and early observation setting (\framework+ and \framework++, respectively) and LSTM-CC as models for comparison; results are reported on 10 of the 43 total subreddits -- 5 showing most response towards observation (top half), and 5 with least response (bottom half). Abbreviations of subreddits follow the same definitions in Table~\ref{tab:subreddit_aggregate}.}
\small
    \centering
    \begin{tabular}{|l|ccccc|}
    \hline
{\bf Model} & AR & Anime & FF & EI & UPO \\
\hline
\framework+ & 37.184 & 37.545 & 39.060 & 38.219 & 37.011 \\
\framework++ & 26.031 & 26.004 & 27.419 & 27.001 & 28.037 \\
LSTM-CC  & 110.34  & 108.69  & 129.078  & 121.336  & 115.414  \\
\hline
{\bf Model} & india & soccer & NBB & business & TNF \\
\hline
\framework+ & 34.323 & 31.095 & 31.866 & 30.519 & 33.911 \\
\framework++ & 27.009 & 24.753 & 24.146 & 26.001  & 25.686\\
LSTM-CC      & 168.91 & 177.247 & 163.077 & 165.325 & 162.12\\
\hline
    \end{tabular}
    \label{tab:subreddit_observe}
\end{table}

\begin{table}[!t]
\caption{MAPE and Kendall's $\tau$   to predict future chatter using \framework\ and LSTM-CC, each with varying size of the prediction window. \framework-$E_x$ and \framework-$Z_x$ signify \framework\ in $1$ hour early observation and zero-shot setting using a prediction window $x$ days long, respectively.}
\small
    \centering
    \begin{tabular}{|l|c|c|}
    \hline
{\bf Model} & {\bf MAPE} & $\tau$ \\
\hline
\framework-$E_1$   & 19.020 & 0.5145 \\
\framework-$E_{10}$   & 21.979 & 0.4610 \\
\framework-$E_{20}$   & 25.224 & 0.4437 \\
\framework-$E_{30}$   & 26.042 & 0.4361 \\
\hline
\framework-$Z_1$   & 22.198 & 0.4812 \\
\framework-$Z_{10}$   & 25.064 & 0.4334 \\
\framework-$Z_{20}$   & 32.719 & 0.4098 \\
\framework-$Z_{30}$   & 33.142 & 0.4042 \\
\hline
    \end{tabular}
    \label{tab:pred-window}
\end{table}

\paragraph{\bfseries Zero-shot vs. Early Observation}
Much similar to exogenous and endogenous influence signals, the role of early-observation to predict future chatter differs between subreddits. In Table~\ref{tab:subreddit_observe}, we explore this phenomena by comparing \framework\ in zero-shot and early observation settings. We also gauge the performance of LSTM-CC as this ablation variant depends purely on comment arrival in observation window to predict future chatter. For the subreddits in the top half of Table~\ref{tab:subreddit_observe} the performance gain with moving from zero-shot to early observation regime is significantly higher compared to those in the bottom half. LSTM-CC follows the same pattern, with MAPE for each of the top-half subreddits being substantially lower than the global average (see Table~\ref{tab:result_overall_agg}) and higher for the top-half subreddits. Various characteristics of the subreddits might be put as responsible for this: size of the subreddit (large subreddits like AskReddit embody complex dynamics of user interests, resulting in chatter signals that can not be modeled using influence signals alone), small secluded subreddits like EcoInternet with focus of discussion not much available over news or rest of the Reddit, etc.

\section{Related Work}
\label{sec:Related}

Most prior work on engagement prediction requires knowledge of the underlying social network.  Such systems are mostly based on modeling information or influence diffusion at a microscopic level
\citep{kupavskii2012retweetcascade, cheng2014cascadepred}, recently enhanced with point processes \citep{zhao2015seismic, kobayashi2016tideh}.  Separation of exogenous and endogenous influences \citep{de2018endoexo} in the predictive model can increase interpretability and accuracy.  Some systems predict if, after observing $k$ instances of influence or transfer along edges, the process will cascade to over $2k$ transfers `eventually'~\citep{cheng2014cascadepred}.  Most such systems use a richly designed space of temporal, structural, and contextual features, along with the standard supervised classifiers, to predict the evolving dynamics of the cascades.

In contrast, we do not assume any knowledge of an underlying  network. Moreover, we are not tracking the diffusion of any uniquely identifiable content like an image or a hashtag.  There is no one-to-one mapping between a specific news story and a network community.  In fact, the same event may be reported in multiple news stories. While subrredits generally have overlapping interests, the extent of exogenous influence of a news story within a subreddit is topical in nature. Endogenous influence within a community depends on the invisible and possibly transient social links.

\citet{guerini2011virality} explore the prediction of viral propagation, based solely on the textual features and not the social network structure, which is closer in spirit to \framework\ compared to network-assisted prediction. They track the spread of specific identified information items (short `stories').  A story can be submitted only once, unlike multiple submissions on a topic in our setting. They propose hardwired definitions of appreciation, discussion, controversiality, and `white' and `black' buzz, then use an SVM classifier to predict such labels successfully.  \citet{aswani2017socialbuzz} presented similar studies.  \citet{shulman2016predictability} found early adoption a stronger predictor of later popularity than other content features.  \citet{weng2012competition} showed how limited attention of individuals causes competition in the evolution of memes.

\citet{peng2018emerging} seek to predict (as early as possible) emerging discussions about products on social media without information about the social network structure.  Like \citet{guerini2011virality}, they engineer a variety of rich features, including author diversity, author engagement, competition from other products, and temporal, content, and user features in a conventional classifier.  Their task is thus limited to a vertical domain (products), and the prediction task is discrete classification (will a burst of activity emerge or not).  In contrast, we seek to predict quantitative levels of chatter. Unlike both \citet{peng2018emerging} and \citet{guerini2011virality}, we avoid extensive feature engineering and instead focus on the design of a deep network that integrates exogenous and endogenous influences.

Chatter intensity or related quantities have been used for predicting other social outcomes as well.  \citet{asur2010predicting} used a   linear regression  
to predict their box-office success.
Other examples such as election outcome and stock movements are surveyed by \citet{yu2012socialprediction}.  Thus, our work on chatter intensity prediction opens up avenues toward such compelling downstream applications.

\section{Conclusion}
\label{sec:End}

Activity prediction usually depends on knowledge of the underlying social network structure.  However, on several important social platforms, the social network is incomplete, not directly observable, or even transient. We introduce the problem of predicting social chatter level without graph information, and present a new deep architecture, \framework, for this setting.  \framework\ combines deep text representation with a recurrent network that tracks the temporal evolution of the state of a community with latent connectivity.  Without knowledge of social network topology, \framework\ achieves new state-of-the-art accuracy.  Here we have regarded chatter as caused by news events but not vice versa, whereas chatter is having increasing effects on the real world.  Modeling such feedback effects may be a natural avenue for future work.

\bibliographystyle{ACM-Reference-Format}
\bibliography{ref}

\newpage
\appendix
\thispagestyle{plain}
\makeatletter
\twocolumn[\centering \@titlefont \ztitle \\
\vspace{.6ex}
\LARGE (Appendix / Supplementary Material) \par \bigskip
\makeatother ]

\section{Corpus Preprocessing}
\label{appendix:preprocess}
We use same strategy for text cleaning of both news articles and submissions. After tokenization, replacing URLs, and converting numeric values to their textual counterpart, we set a maximum document frequency of 0.8 (fraction of the total number of news articles and submissions) and minimum document frequency of $5$ (absolute count) to exclude stopwords and extremely rare words. We trained the Word2Vec model for $500$ iterations with window size set to $10$ and output dimension $100$. We take the maximum length of texts to be $50$ and $100$ words for submissions and news articles, respectively.

\section{Design details of \framework}
\label{appendix:parameters}
For each of the $43$ subreddits represented as one-hot vector, the subreddit embedding layer outputs a $32$-dimensional vector. Every weight matrix is randomly initialized using Xavier initialization.  All bias matrices are initialized with zeros.

\subsection{Organization of Convolution Blocks}
\label{appendix:conv}

\begin{figure}[h]
\centering
\includegraphics[width=\columnwidth]{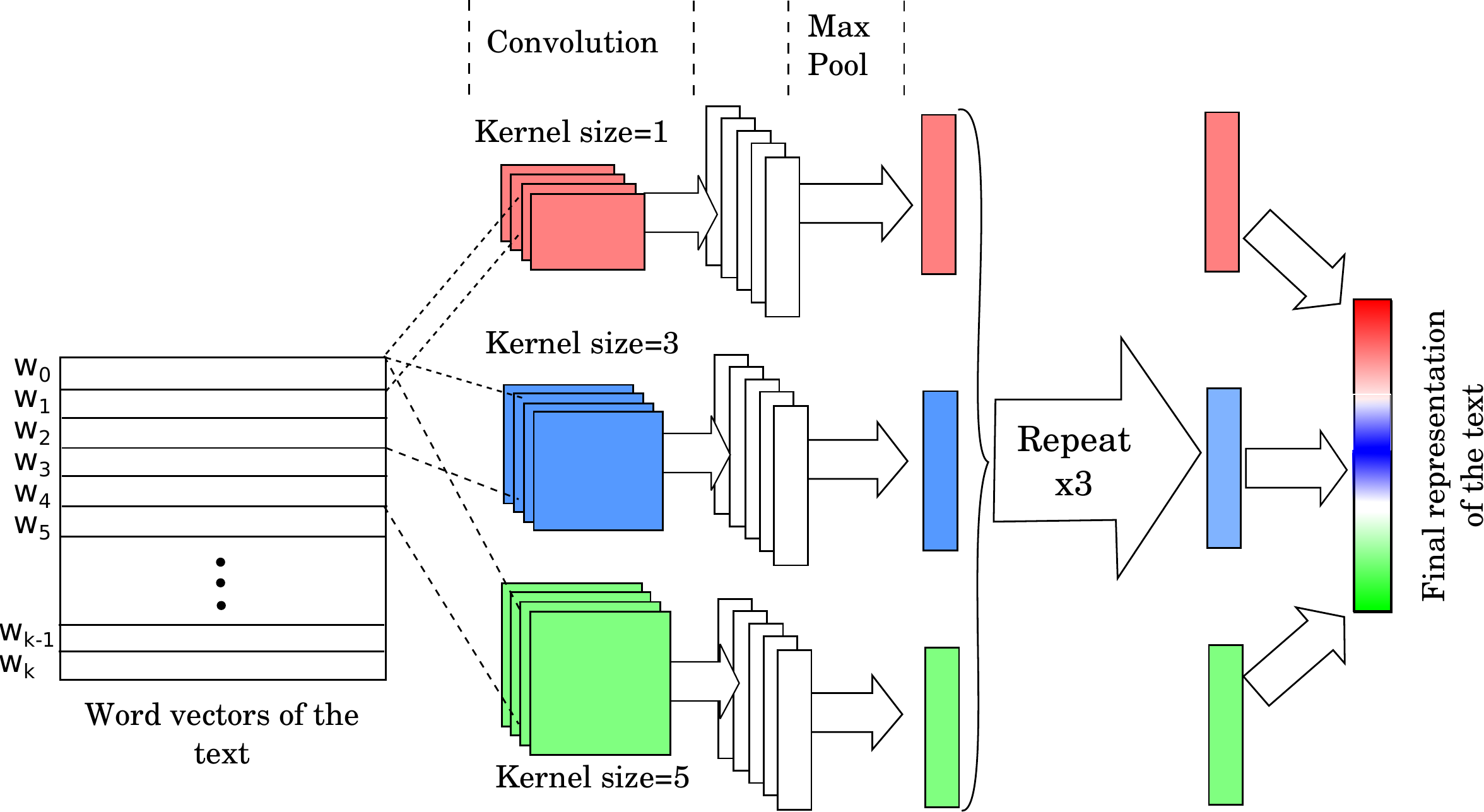}
\caption{Static convolution of news and submission texts to obtain the latent feature representations. The three parallel branches of convolution-maxpooling is repeated thrice, and then concatenated to produce the feature map. The initial list of the word vectors comes from the word embedding layer.}
\label{fig:text_cnn}
\end{figure}

\begin{figure}[h]
\centering
\includegraphics[width=\columnwidth]{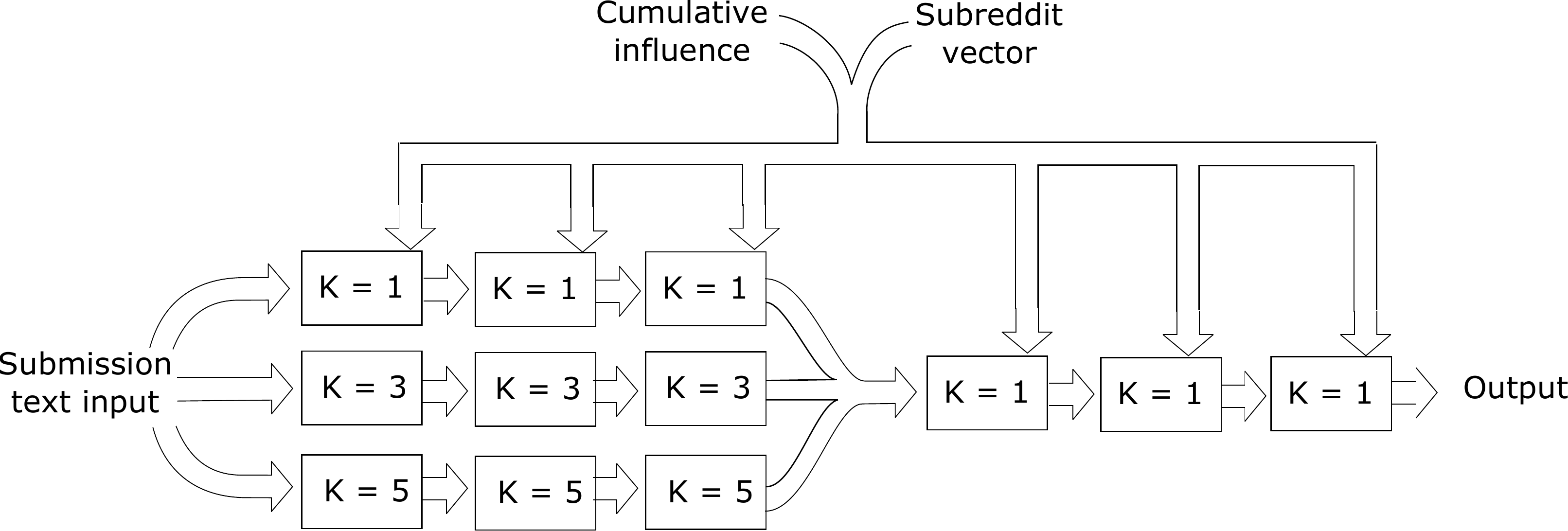}
\caption{The time-evolving convolution component. $K$~denotes size of convolution filters. Cumulative influence and subreddit vectors corresponding to the input submissions are the control inputs for every convolution layer. }
\label{fig:ada_cnn}
\end{figure}

\framework\ uses two separate stackings of convolution-maxpool operations: static convolution block and time-evolving convolution block (components $1$ and $3$ in Figure~\ref{fig:whole_model}). Internal organizations of the blocks are shown in Figure~\ref{fig:text_cnn} (static) and Figure~\ref{fig:ada_cnn} (time-evolving). For all the convolution operations in the static block and the branched segments of the time-evolving block we use padding to keep the size of output feature maps to be same as inputs. For the last three convolution operations in the temporal block we do not use any padding (as the kernel size is 1).

All the branches of convolution-maxpooling in both the blocks have number of filters $128$, $64$, and $32$, successively. Last three convolution operations in time-evolving block have filter numbers $64$, $32$, and $1$, successively.

\subsection{Parameters of Recurrent Units}
\label{appendix:recurrent}

The news-aggregating and the submission-aggregating GRUs (see component (2) in Figure~\ref{fig:whole_model}) both have hidden state size equal to $128$. The LSTM layer aggregating comment arrivals in the early observation window uses hidden state of size of $8$.

\section{\framework{} training details}
\label{appendix:training}

To initialize the word embedding layers, we train a skip-gram Word2Vec model on the training split of the news and submission data. The resulting word vectors are of size $100$, and they are further trained while training \framework.  \framework\ is optimized using Adam optimizer~\citep{kingma2014adam}, with the learning rate set to $0.00001$. As the model works in an online setting, the batch size is set to one. While training \framework, we reset the states of stateful GRUs after every epoch (after testing on the validation data).

\framework\ is trained on a Intel Xeon Processor ($16$ cores, $32$ GB RAM) with NVIDIA Quadro K6000 GPU. Each training iteration of full \framework\ takes 13 hours and 22 minutes (roughly). We trained all the models for $25$ iterations and save top $5$ best models (on validation loss). All the results reported are averaged over these $5$ models.

\end{document}